\documentclass{article}
\usepackage{srcltx}
\usepackage{amsmath}
\usepackage{amssymb}
\usepackage{amsfonts}
\usepackage{latexsym}
\usepackage{textcomp}
\usepackage{appendix}
\usepackage{multirow}
\usepackage{booktabs}
\usepackage{url}
\usepackage{array}
\usepackage{marvosym}
\usepackage{graphics}
\usepackage{graphicx}
\usepackage[table]{xcolor}
\usepackage{epsfig}
\usepackage{float}
\usepackage{authblk}

\usepackage[caption=false]{subfig}
\newcommand{\overbar}[1]{\mkern 1.5mu\overline{\mkern-1.5mu#1\mkern-1.5mu}\mkern 1.5mu}

\title{An analysis of cryptocurrencies conditional cross correlations}

\author[1]{Nektarios Aslanidis \thanks{nektarios.aslanidis@urv.cat}}
\author[2]{Aurelio F. Bariviera \thanks{aurelio.fernandez@urv.cat}}
\author[1]{Oscar Mart\'inez-Iba\~nez \thanks{oscar.martinez@urv.cat}}
\affil[1]{\scriptsize Universitat Rovira i Virgili, Department d'Economia,  CREIP, Avinguda Universitat 1, Reus 43204, Spain}
\affil[2]{\scriptsize  Universitat Rovira i Virgili, Department of Business, Av. Universitat 1, 43204 Reus, Spain}

\begin{document}
\maketitle

\begin{abstract}
This letter explores the behavior of conditional correlations among main cryptocurrencies, stock and bond indices, and gold, using a generalized DCC class model. From a portfolio management point of view, asset correlation is a key metric in order to construct efficient portfolios. We find that: (i) correlations among cryptocurrencies are positive, albeit varying across time; (ii) correlations with Monero are more stable across time; (iii) correlations between cryptocurrencies and traditional financial assets are negligible. 
\end{abstract}

\section{Introduction}
Almost 10 years ago, triggered by the seminal paper by \cite{Nakamoto}, a new type of financial asset was born. Based on the concept of a distributed ledger, blockchain technology is able to validate operations, without the necessity of a central trusted authority.  The foremost application of blockchain is the validation of financial transactions. As a consequence, several assets (self-called cryptocurrencies) emerged as an alternative to standard fiat money. 

There is no consensus on the 'currency' status of the so-called cryptocurrencies. \cite{Polasik2015} report an increasing number of businesses and organizations accepting payments in bitcoin, and \cite{Kristoufek2015} highlights the usefulness of bitcoin as medium of exchange. However, cryptocurrency volatility raises doubt about their suitability as a store of value. 

Whether they are currencies or just plain financial assets, investors have been increasingly interested in them.  The cryptocurrency econsystem goes beyond Bitcoin. There are more than 1800 coins based on blockchain technology, which are traded in a frenzy, 24/7 market. As of October 2018 the daily traded volume of this market exceeds 10 USD billions  (\cite{coinmarketcap}).

Academics are not unaware of this phenomenon. Scopus database (as of February 2019) includes 1882 documents with 'bitcoin' in its title, abstract or keywords, half of them published in 2017 and 2018. These articles cover different topics of this novel product: legal concerns, economic perspectives or computer pecularities. One of the key aspects of is the number of cryptocurrencies that enter and leave the market. \cite{Elbahrawy2017} find that the average monthly birth and death since 2015 is very similar, meaning approximately seven new cryptocurrencies appear, and a similar figure disappear every week. 

Most previous studies have been focused solely on Bitcoin. Taking into account the variety of crypto-assets, one natural line of research could be to study simultaneously some of them. There are other cryptocurencies based on the same blockchain technology. Consequently, at first sight, could seem essentially the same asset. None of them is endorsed by governments, nor they have physical (paper or metal) support. Additionally, they seem to have a null intrinsic value. Two questions that could be raised are: (i) Do these currencies follow a related underlying process? (ii) is this relationship stable across time?
The first question was (partially) answered by \cite{Bariviera2018}, when studying the 12 most important cryptocurrencies at high frequency sampling. They detect three different underlying dynamics. Most cryptocurrencies in the sample followed the same stochastic process as Bitcoin. However, Ethereum and Ethereum Classic (on one side), and Dash and NEM (on the other side) follow their own stochastic dynamic. 
We aim, in this letter, to reexamine the first question using an alternative methodology, and study the second question. To the best of our knowledge, this is the first study applying dynamical conditional correlation analysis to the cryptocurrency market. 
 
The letter is organized as follows. Section \ref{sec:literature} review some key aspects on cryptocurrency and related literature. Section \ref{sec:dcc} briefly presents the methodology. Section \ref{sec:results} describes data and discusses the main findings. Finally, Section  \ref{sec:conclusions} outlines the conclusion of our analysis.

\section{Brief literature review \label{sec:literature}}
\cite{Urquhart2016} used a set of tests aimed at identifying autocorrelations, unit roots, nonlinearities and long range dependence in Bitcoin returns. The results show evidence of information inefficiency in the Bitcoin market. However, when the author splits the sample into two sub-periods, it is found inefficiency is an issue mainly in the first part of the sample. Later, \cite{Nadarajah2017} reexamines the data using power transformations of daily returns, without rejecting the null hypothesis of informational efficiency. \cite{Bourietal2017} study Bitcoin's return-volatility behavior before and after the severe market crash of 2013, and show evidence of serial autocorrelation in Bitcoin returns. \cite{Bouri20171frl} scrutinize hedge and safe haven properties of Bitcoin \textit{vis-\`a-vis} international stock and bond indices and several currencies. The main finding is the the Bitcoin proves useful as a diversifier rather than as a hedge instrument. Finally, \cite{Balcilar20177em} detect nonlinearities in the return-volume relationship, which allows for return prediction.

Furthermore, \cite{Bariviera2017} documents that the Bitcoin market exhibits time-varying information efficiency and persistence in volatility. The policy implication here is that the market becomes prone to large swings (either positive or negative ones). Taking this into account, 
\cite{DonierBouchaudPlosOne} study different measures of liquidity as early warning signs of bitcoin market crash.

\cite{Dyhrberg2016} studies simultaneously Bitcoin, gold and the dollar, using GARCH models, finding similarities in all three assets. In particular, bitcoin and gold react to the same variables in a GARCH model, and also bitcoin reacts to federal fund rates, as in the case of a fiat currency. 

One key aspect in portfolio theory, and broadly in financial economics, is the correct assessment of correlation returns among different assets. Such metric has important implications regarding portfolio construction, risk analysis and hedging. \cite{Corbet2018} employ the generalized variance decomposition methodology by \cite{Diebold2012}. They find that three major cryptocurencies (Bitcoin, Ripple, Litecoin) are rather isolated from other assets such as gold, stock or bond indices, offering diversification opportunities to investors. 

Given the burgeoning literature on this topic, we refer to \cite{Corbet2018b} and \cite{SmithKumar}  for excellent reviews of the empirical literature. 

\section{Time-varying conditional correlations \label{sec:dcc}}

Let $r_t$ denote an $N$-dimensional vector time series (zero-mean asset returns) with time-varying conditional covariance matrix:
\begin{equation}
   Var[r_t|\Im_{t-1}]=E[r_t r^{'}_t|\Im_{t-1}]=H_t \qquad t=1,\dots,T
\label{eq:variance}
\end{equation}

where $\Im_{t-1}$ is the information set at time $t$. The conditional covariance matrix can be decomposed as (see, \cite{Engle2002}, among others): 
\begin{equation}
  H_t = D_t R_t D_t
\label{eq:variance_decomposition}
\end{equation} 
where $ D_t \equiv diag( \sqrt{h_{1,t}},\dots,\sqrt{h_{N,t}})$ is a diagonal matrix with the square root of the conditional variances on the diagonal. The matrix $R_t$, with the $(i,j)$-th element denoted as $\rho_{ij, t}$, is the possibly time-varying correlation matrix with $\rho_{ii, t}=1$, $j = 1,\dots,N$ and $t = 1,\dots,T$. The standardised returns are denoted by $\varepsilon_t=D^{-1}_t r_t=(\varepsilon_{1t},\dots,\varepsilon_{Nt})'$ .

One of the most frequently used methodologies in capturing the time-varying structure of the correlations is the Dynamic Conditional Correlation (DCC) model which assumes that the conditional correlation evolves linearly according to a simple GARCH (1,1)-type structure (\cite{Engle2002}). The DCC framework has become popular among academics and practitioners. 

In a multivariate framework, the basic DCC may be too restrictive. In particular, this model implies that all correlations pairs have the same response to news and decay parameters. For our application (consisting of 4 assets) we adopt a flexible generalization of the DCC, which allows for correlation-specific news parameters, while the decay parameter is assumed to be the same across correlation pairs, in order to keep the model tractable. This generalized DCC model, studied initially in \cite{Cappiello2006} and \cite{Hafner2006} and later in \cite{Aslanidis2013}, is given by:
	\begin{equation}
Q_{t}=(\overbar{Q}-A' \overbar{Q} A -\beta \overbar{Q}) + A' \varepsilon_{t-1} \varepsilon^{'}_{t-1} A+ \beta Q_{t-1}
\label{eq:generalDCC}
\end{equation} 

\begin{equation}
   R_t={Q^*}^{-1}_t\;Q_t\;{Q^*}^{-1}_t
\label{eq:generalDCC2}
\end{equation}
where $A\equiv diag(\alpha_1,\dots,\alpha_N)$ is parameter diagonal matrix (the implied news parameters are $\alpha_i \alpha_j$ for $i\neq j$ ) and $\beta$ is the decay parameter. As usual, we rescale the quantity $Q_t$ in Eq. \ref{eq:generalDCC2} to obtain a proper correlation matrix, with $Q^*_t$  being a diagonal matrix composed of the square roots of the diagonal elements of  $Q_t$ .
So, the basic DCC is obtained as a special case of the generalized DCC if the matrix $A$ is replaced by the scalar $\alpha$.

\section{Data and Results \label{sec:results}}

We use daily  price data of four cryptocurrencies, and three traditional financial assets. Cryptocurrencies included in our sample are: Bitcoin (BTC), Dash (DASH), Monero (XMR), and Ripple (XRP). The three traditional assets selected are Standard \& Poors 500 Composite (SP500), S\&P US Treasury bond 7-10Y index (BOND), and Gold Bullion LBM (GOLD). Cryptocurrency data were obtained from https://coinmarketcap.com/, and the other assets data were downloaded from Eikon Thomsom Reuters. The period under examination goes from 21/05/2014 to 27/09/2018. Cryptocurrencies are traded 24 hours a day, 7 days a week. However, traditional assets are traded in organized markets, that are open only during the working week. Therefore we have 1560 observations for cryptocurrencies and 1135 observations for the other assets. 

We show in Table \ref{tab:dailyreturns} the descriptive statistics of daily logarithmic returns of both cryptocurrencies and traditional assets. Considering that both types of assets differ in trading hours, we reduced the dataset of cryptocurrencies to match the number of observations and dates of the traditional assets. Therefore, in this table we only considered data from Monday through Friday, and we computed the logarithmic return of two consecutive observations. 
We would like to highlight the enormous difference between the two assets classes. Mean and standard deviation in cryptocurrencies are between 6 to 144 times larger than traditional assets. 

\begin{table}[htbp]
  \centering
  \caption{Descriptive statistics of daily returns}
 \resizebox{\columnwidth}{!}{   \begin{tabular}{lrrrrrrr}
    \toprule
          & \multicolumn{1}{c}{GOLD} & \multicolumn{1}{c}{S\&P500} & \multicolumn{1}{c}{BOND} & \multicolumn{1}{c}{BTC} & \multicolumn{1}{c}{XRP} & \multicolumn{1}{c}{DASH} & \multicolumn{1}{c}{XRM} \\
    \midrule
    Observations & 1134  & 1134  & 1134  & 1134  & 1134  & 1134  & 1134 \\
    Mean  & -0.0079 & 0.0377 & -0.0031 & 0.2221 & 0.4506 & 0.2351 & 0.3166 \\
    Median & 0.0000 & 0.0268 & 0.0000 & 0.2386 & -0.2181 & -0.0722 & 0.0622 \\
    Std. Dev. & 0.8104 & 0.7744 & 0.3216 & 4.5703 & 8.8985 & 7.9748 & 8.6688 \\
    Min   & -3.2239 & -4.1843 & -1.5377 & -26.4311 & -57.0455 & -73.3201 & -36.6830 \\
    Max   & 4.6184 & 3.8291 & 1.3196 & 27.8435 & 136.3081 & 50.0787 & 69.1884 \\
    Skewness & 0.2312 & -0.5831 & -0.0559 & -0.1335 & 3.9416 & -0.1229 & 1.3535 \\
    Kurtosis & 5.6593 & 6.9085 & 4.0056 & 8.4510 & 57.5688 & 13.6400 & 12.0729 \\
    Jarque Bera & 344.2570 & 786.0706 & 48.3706 & 1407.3155 & 143635.1882 & 5352.0396 & 4235.7217 \\
    Probability & 0.0000 & 0.0000 & 0.0000 & 0.0000 & 0.0000 & 0.0000 & 0.0000 \\
    \bottomrule
    \end{tabular}%
		}
  \label{tab:dailyreturns}%
\end{table}%

\subsection{Cryptocurrencies market}

\begin{figure}
\centering
\subfloat[]{\includegraphics[width = 0.4\textwidth]{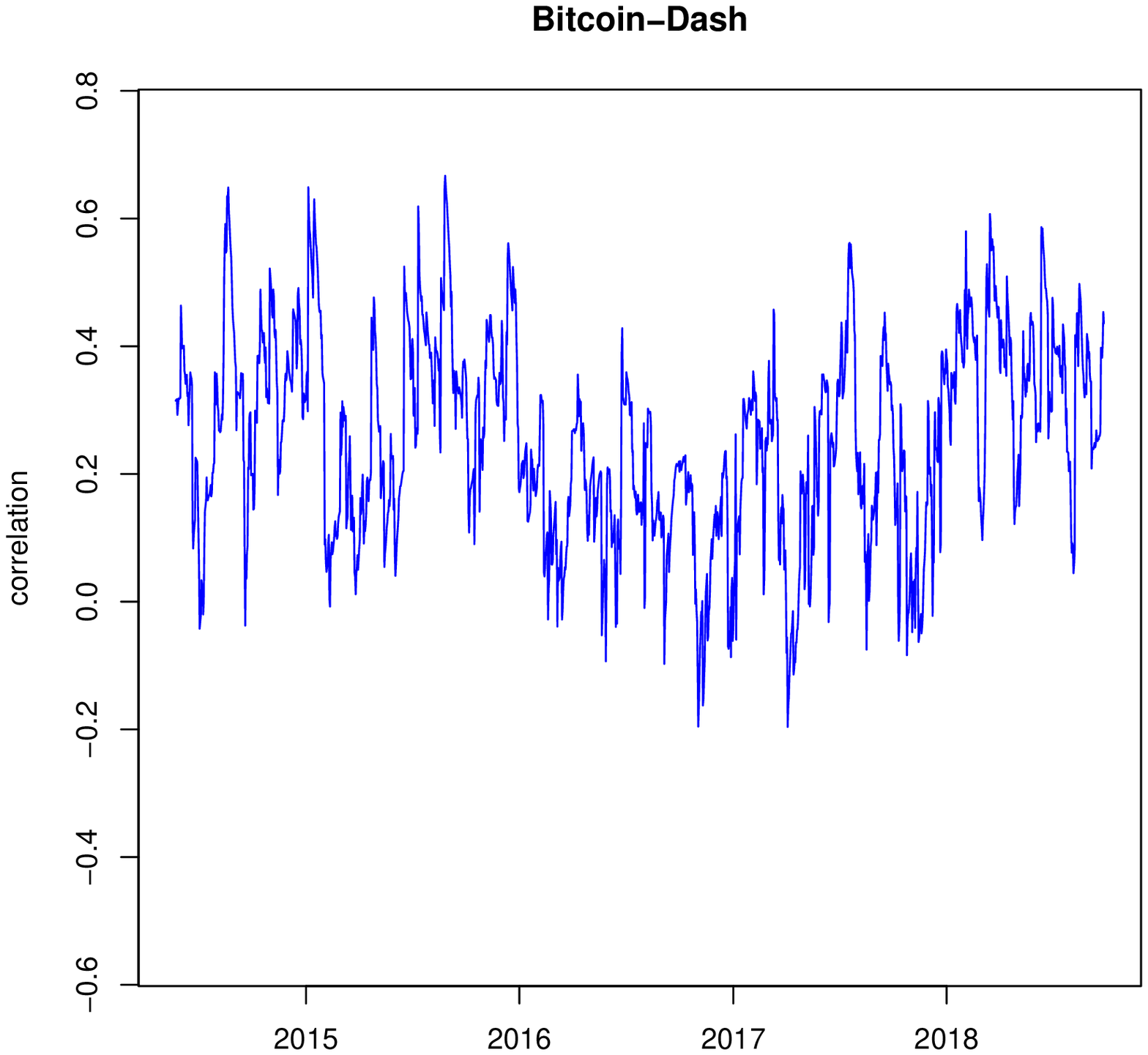}} 
\subfloat[]{\includegraphics[width = 0.4\textwidth]{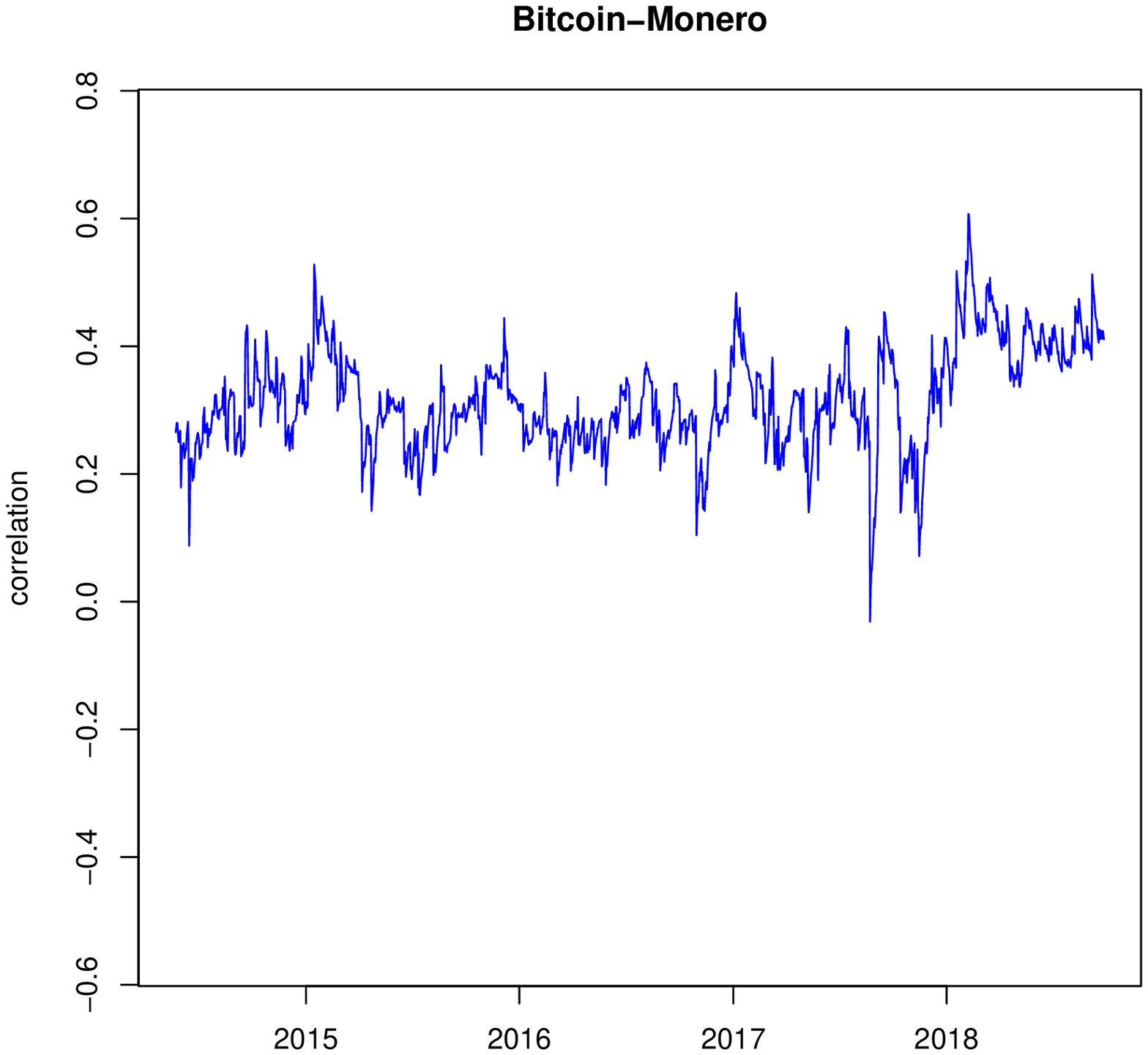}}\\
\subfloat[]{\includegraphics[width = 0.4\textwidth]{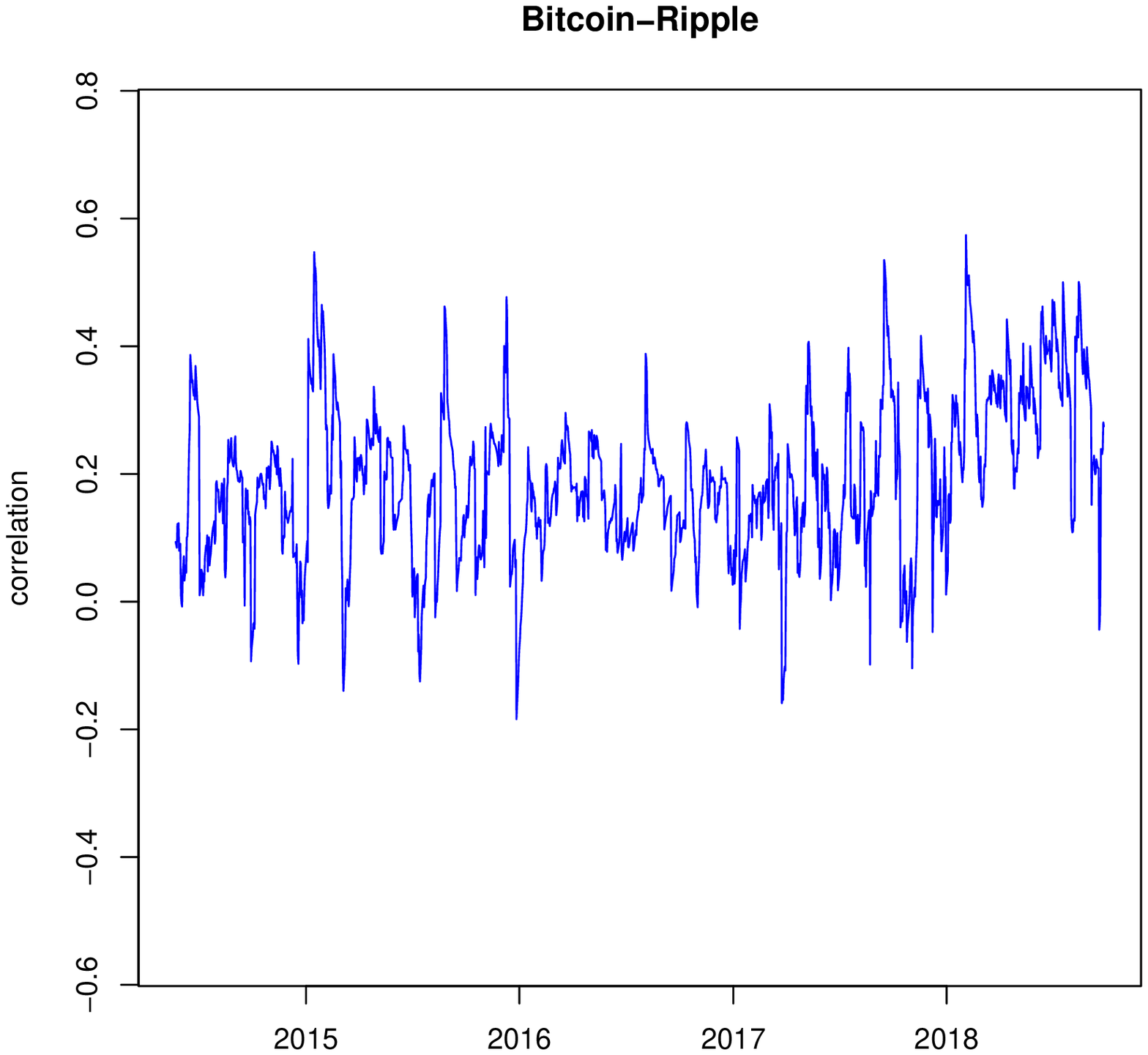}}
\subfloat[]{\includegraphics[width = 0.4\textwidth]{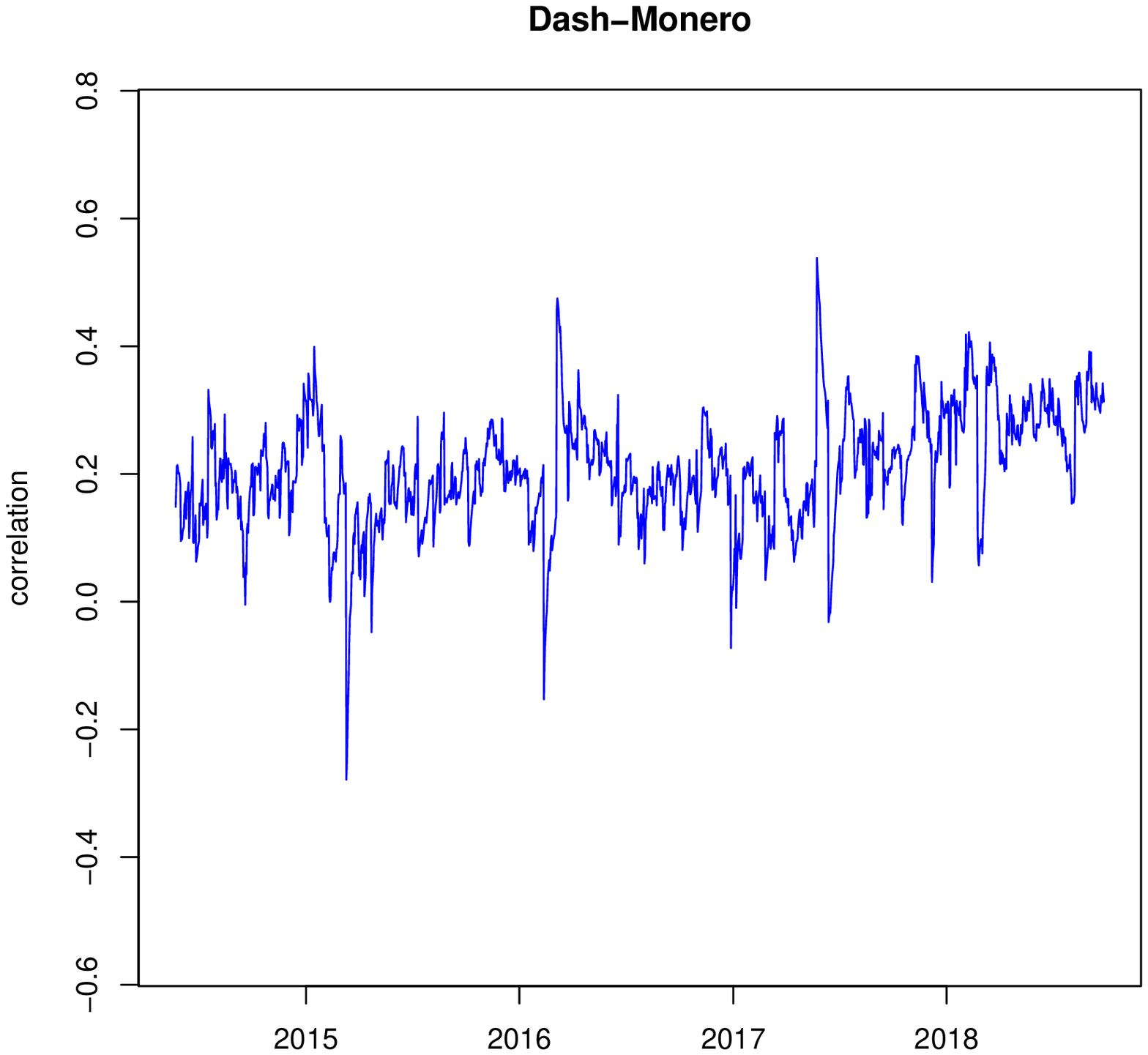}} \\
\subfloat[]{\includegraphics[width = 0.4\textwidth]{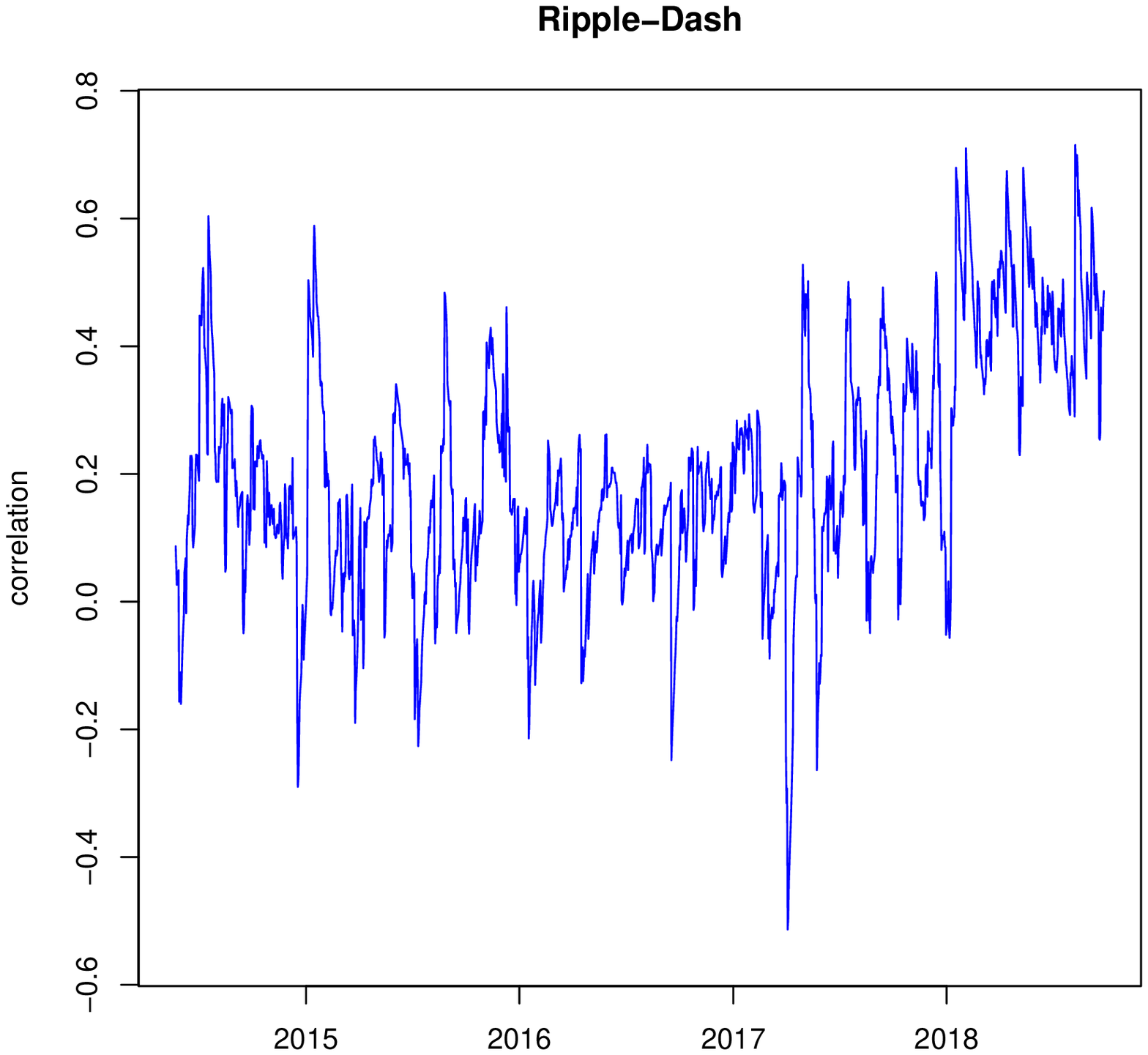}} 
\subfloat[]{\includegraphics[width = 0.4\textwidth]{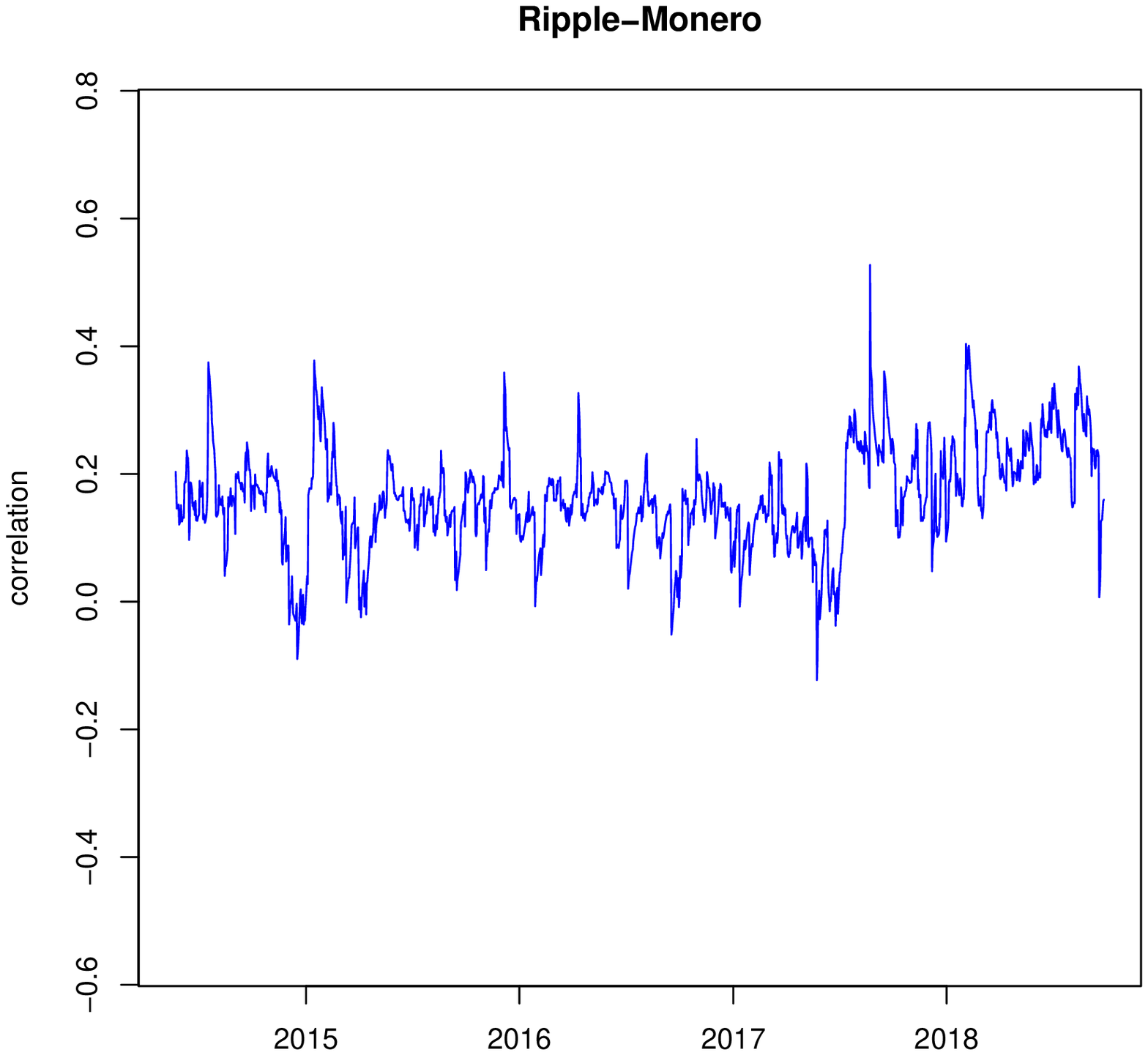}} 
\caption{Cryptocurrencies pairwise correlations}
\label{fig:correlations}
\end{figure}

We compute the dynamic cross correlations, according to Generalized DCC, considering the full dataset. We recall that cryptocurrencies are traded 24/7. Thus, there are no days without trading or weekends. Correlation results for each correlation pair is displayed in Figure \ref{fig:correlations}. 

Our study detects several interesting features in the cryptocurrency ecosystem. Table \ref{tab:descriptive} displays descriptive statistics of estimated correlations, with the model estimates being $A=diag(0.233, 0.239, 0.289, 0.133)$ and $\beta=0.866$. We find that, on average, correlation between cryptocurrencies is positive. 
Note there is low variablility in all correlation pairs with Monero. This can be explained by its low alpha parameter, $\alpha_4=0.133$. So the implied news parameters for these three correlations pairs are $0.233*0.133=0.031$, $0.239*0.133=0.032$ and $0.289*0.133=0.038$ for  Bitcoin-Monero, Ripple-Monero and Dash-Monero, respectively.
The most variable correlation is found between Ripple and Dash, whose correlation spans from  $+0.71$ to $-0.51$.

\begin{table}[htbp]
  \centering
\caption{Descriptive statistics of dynamical cross correlation of cryptocurrencies' pairs}
  \resizebox{\columnwidth}{!}{
		\begin{tabular}{lrrrrrr}
    \toprule
          & \multicolumn{1}{c}{BTC} & \multicolumn{1}{c}{BTC} & \multicolumn{1}{c}{BTC} & \multicolumn{1}{c}{XRP} & \multicolumn{1}{c}{XRP} & \multicolumn{1}{c}{DASH} \\
          & \multicolumn{1}{c}{XRP} & \multicolumn{1}{c}{DASH} & \multicolumn{1}{c}{XMR} & \multicolumn{1}{c}{DASH} & \multicolumn{1}{c}{XMR} & \multicolumn{1}{c}{XMR} \\
    \midrule
    Mean  & 0.1912 & 0.2535 & 0.3161 & 0.2035 & 0.1639 & 0.2072 \\
    Median & 0.1856 & 0.2586 & 0.3073 & 0.1796 & 0.1597 & 0.2065 \\
    Maximum & 0.5740 & 0.6669 & 0.6070 & 0.7151 & 0.5272 & 0.5384 \\
    Minimum & -0.1840 & -0.1962 & -0.0312 & -0.5134 & -0.1227 & -0.2789 \\
    Std. Dev & 0.1216 & 0.1579 & 0.0793 & 0.1862 & 0.0792 & 0.0896 \\
    Skewness & 0.1690 & -0.0581 & 0.1220 & 0.1240 & 0.0379 & -0.2694 \\
    Kurtosis & 3.1208 & 2.5690 & 3.5280 & 3.1768 & 3.6295 & 4.6496 \\
    Jarque-Bera & 8.3777 & 12.9523 & 21.9910 & 6.0297 & 26.1341 & 195.7463 \\
    Probability & 0.0152 & 0.0015 & 0.0000 & 0.0491 & 0.0000 & 0.0000 \\
    \bottomrule
    \end{tabular}%
		}
  \label{tab:descriptive}%
\end{table}%

Our results have important implications for portfolio analysis. An investor, who constructs a dynamic hedge portfolio of cryptocurrencies could take into account the stable correlations involving Monero.

\subsection{Traditional assets markets}
\begin{figure}
\centering
\subfloat[]{\includegraphics[width = 0.4\textwidth]{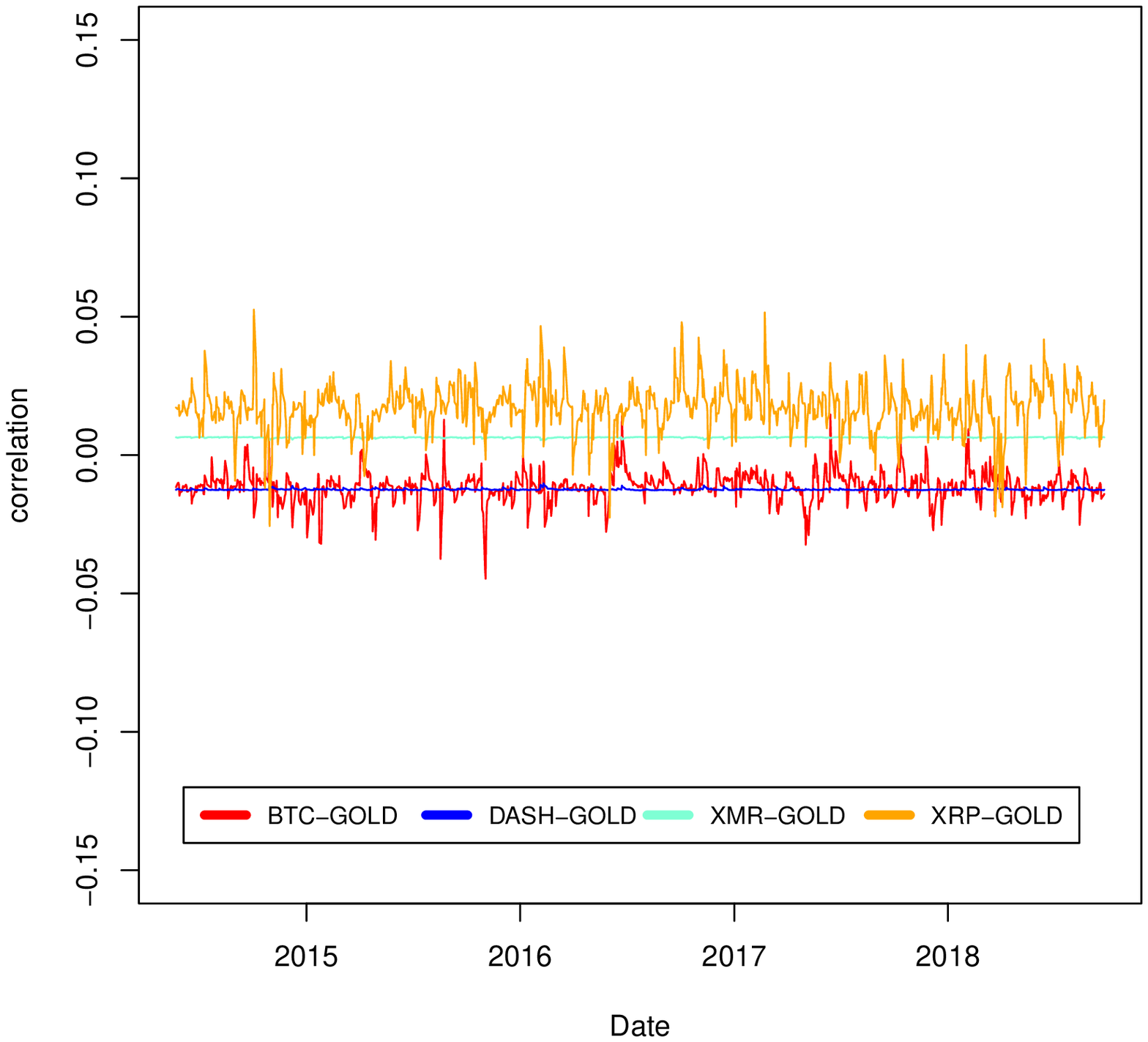}} \\
\subfloat[]{\includegraphics[width = 0.4\textwidth]{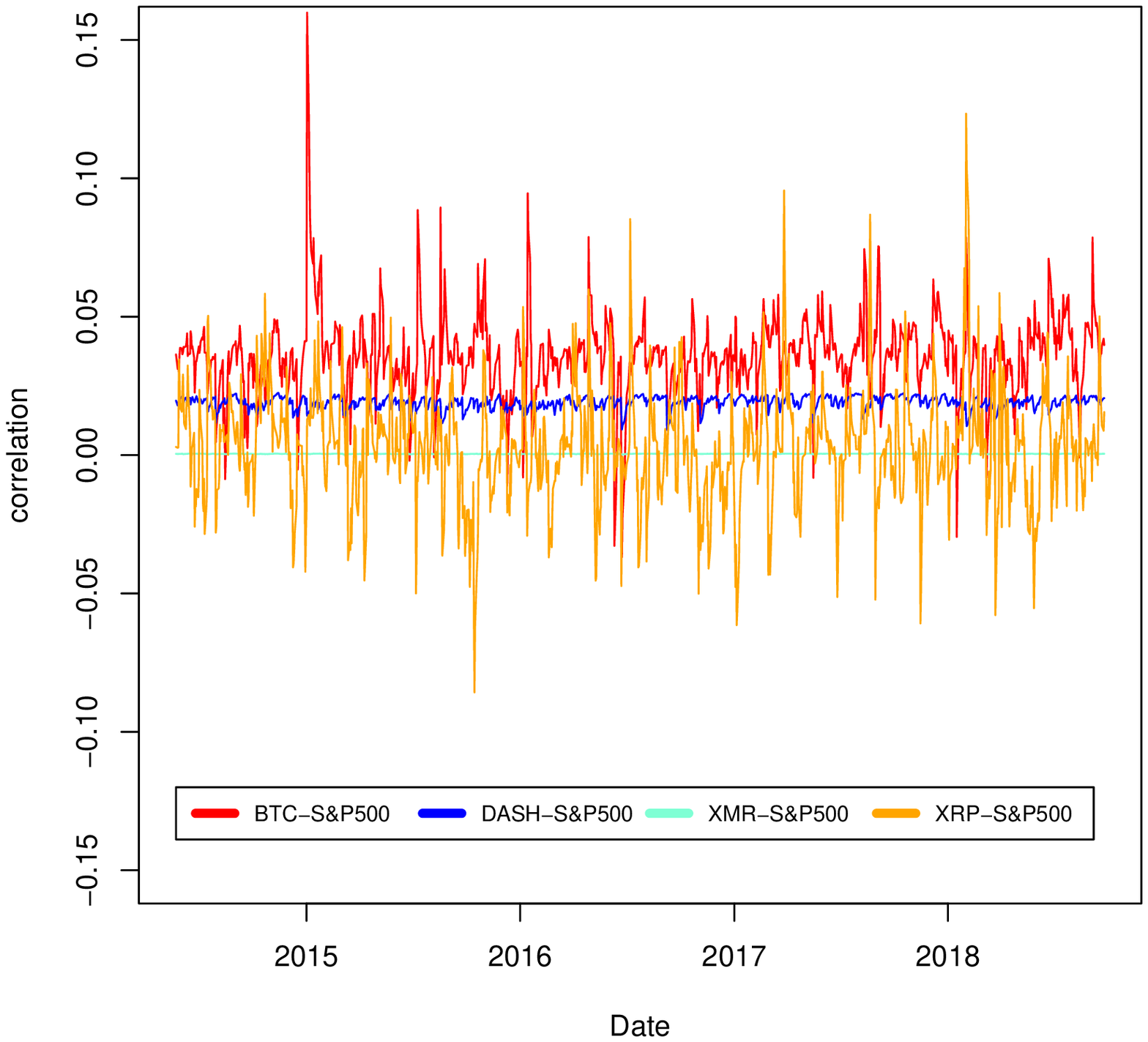}}\\
\subfloat[]{\includegraphics[width = 0.4\textwidth]{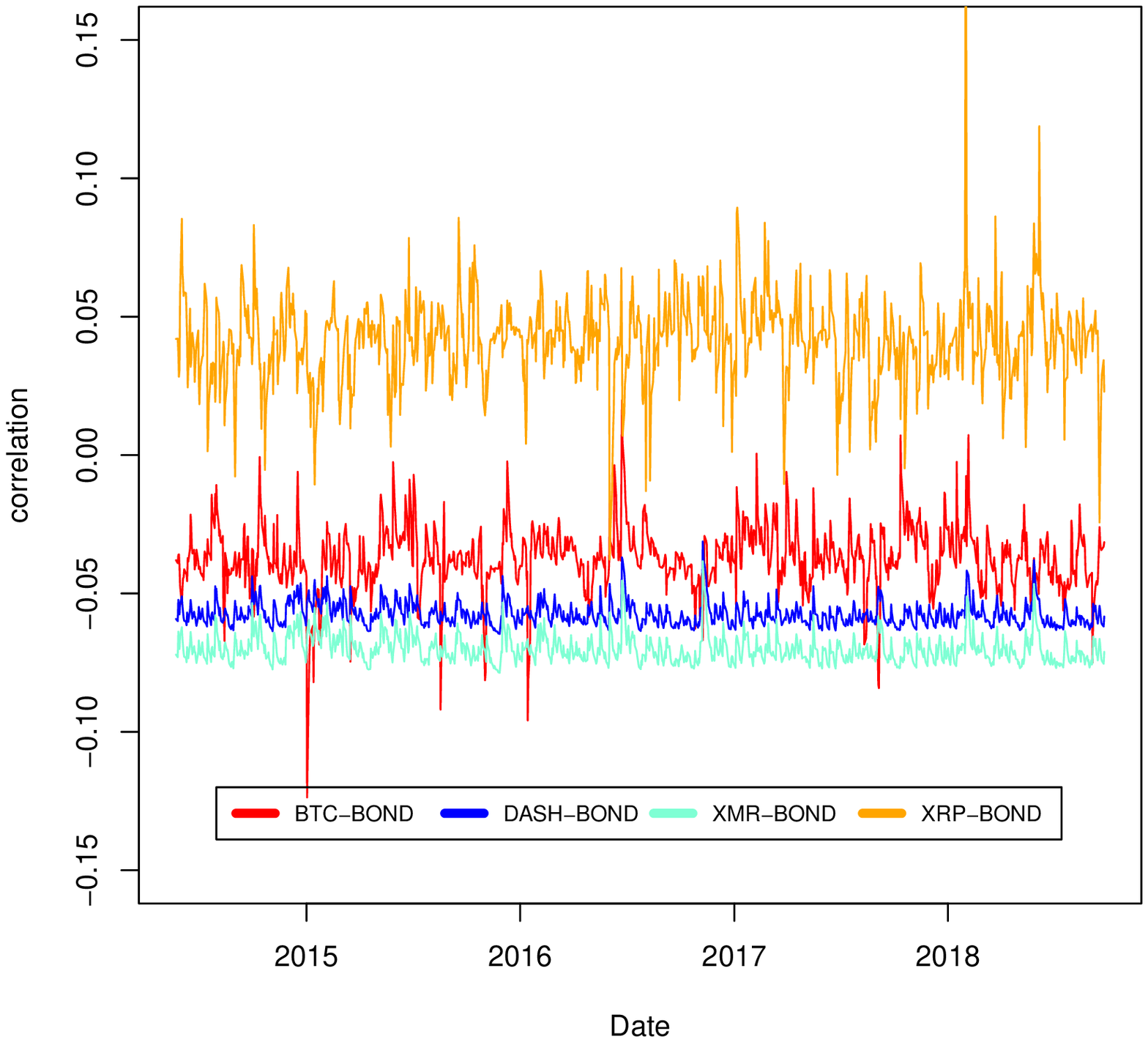}}
\caption{Cryptocurrencies and traditional assets pairwise correlations}
\label{fig:correlations2}
\end{figure}

In this subsection we compute cryptocurrencies cross correlations against traditional assets such as Standard \& Poors 500 Composite (SP500), S\&P US Treasury bond 7-10Y index (BOND), and Gold Bullion LBM (GOLD). Taking into account that the latter assets only trade Monday through Friday, we adapt the cryptocurrency sample to this restriction. Descriptive statistics of these cross-correlations are displayed in Table \ref{tab:corr-trad} and in Figure \ref{fig:correlations2}.

\begin{table*}[htbp]
  \centering
\caption{Descriptive statistics of dynamical cross correlation for cryptocurrency-traditional asset pairs}
  \resizebox{\textwidth}{!}{
			\begin{tabular}{lrrrrrrrrrrrr}
    \toprule
          & \multicolumn{1}{c}{BTC} & \multicolumn{1}{c}{BTC} & \multicolumn{1}{c}{BTC} & \multicolumn{1}{c}{DASH} & \multicolumn{1}{c}{DASH} & \multicolumn{1}{c}{DASH} & \multicolumn{1}{c}{XMR} & \multicolumn{1}{c}{XMR} & \multicolumn{1}{c}{XMR} & \multicolumn{1}{c}{XRP} & \multicolumn{1}{c}{XRP} & \multicolumn{1}{c}{XRP} \\
	        & \multicolumn{1}{c}{GOLD} & \multicolumn{1}{c}{S\&P500} & \multicolumn{1}{c}{BOND} & \multicolumn{1}{c}{GOLD} & \multicolumn{1}{c}{S\&P500} & \multicolumn{1}{c}{BOND} & \multicolumn{1}{c}{GOLD} & \multicolumn{1}{c}{S\&P500} & \multicolumn{1}{c}{BOND} & \multicolumn{1}{c}{GOLD} & \multicolumn{1}{c}{S\&P500} & \multicolumn{1}{c}{BOND} \\
    \midrule
    Observations & 1133  & 1133  & 1133  & 1133  & 1133  & 1133  & 1133  & 1133  & 1133  & 1133  & 1133  & 1133 \\
    Mean  & -0.0114 & 0.0356 & -0.0371 & -0.0124 & 0.0190 & -0.0574 & 0.0064 & 0.0004 & -0.0699 & 0.0171 & 0.0029 & 0.0411 \\
    Median & -0.0113 & 0.0358 & -0.0374 & -0.0125 & 0.0194 & -0.0583 & 0.0064 & 0.0005 & -0.0710 & 0.0172 & 0.0032 & 0.0420 \\
    Std. Dev. & 0.0054 & 0.0145 & 0.0117 & 0.0003 & 0.0022 & 0.0042 & 0.0002 & 0.0001 & 0.0051 & 0.0084 & 0.0204 & 0.0160 \\
    Min   & -0.0446 & -0.0368 & -0.1237 & -0.0128 & 0.0091 & -0.0646 & 0.0052 & -0.0001 & -0.0787 & -0.0256 & -0.0857 & -0.0377 \\
    Max   & 0.0147 & 0.1599 & 0.0198 & -0.0102 & 0.0225 & -0.0312 & 0.0066 & 0.0006 & -0.0384 & 0.0525 & 0.1233 & 0.1748 \\
    Skewness & -0.4068 & 0.5391 & -0.5775 & 2.5126 & -1.0930 & 1.4827 & -2.4022 & -1.4505 & 1.4843 & -0.3647 & 0.3775 & 0.5953 \\
    Kurtosis & 7.9148 & 11.7318 & 8.7038 & 13.2251 & 4.5691 & 6.8779 & 11.9654 & 8.6414 & 6.8898 & 5.9921 & 6.2472 & 12.5905 \\
    Jarque Bera & 1171.6081 & 3654.2501 & 1598.8186 & 6127.8613 & 341.8345 & 1125.0357 & 4884.2173 & 1899.7223 & 1130.3363 & 447.7655 & 524.6903 & 4409.0356 \\
    \bottomrule
    \end{tabular}%
		}
  \label{tab:corr-trad}%
\end{table*}%

We summarize our findings in the correlation matrix displayed in Table \ref{tab:matrix}. We detect three correlation groups. The first group is a well known correlation stocks, bonds and commodities. The second group correspond to cryptocurrencies cross correlations. We detect that this market is positively correlated. Finally, the third group correspond to very low correlations between one cryptocurrency and one traditional asset. We verify, using an alternative methodology, the findings by \cite{Corbet2018}, in the sense that the cryptocurrencies returns are not connected to main financial markets. We also detect that, correlation of traditional assets against Monero are even closer to zero than against other cryptocurrencies. This result could mean that Monero behavior is different from other cryptocurrencies, in the same way as we detected previously that cryptocurrency correlations against Monero are more stable through time. 

However, we believe that this market could not be attractive for diversification pursposes. The reason is that, cryptocurrencies' mean return and stardard deviation are between 1 and 2 orders of magnitude larger than the other traditional assets. As a consequence, a small portion of cryptocurrency, will dominate the stochastic dynamic of the whole portfolio.

\begin{table}[htbp]
  \centering
	  \caption{Correlation matrix. Cross correlation for cryptocurrencies' pairs are computed for the full sample. Cross correlation for cryptocurrency-traditional asset pairs are computed for the reduced sample, corresponding to stock and bonds trading week.}
		\resizebox{\columnwidth}{!}{
    \begin{tabular}{lrrrrrrr}
          & \multicolumn{1}{c}{BTC} & \multicolumn{1}{c}{DASH} & \multicolumn{1}{c}{XMR} & \multicolumn{1}{c}{XRP} & \multicolumn{1}{c}{GOLD} & \multicolumn{1}{c}{S\&P500} & \multicolumn{1}{c}{BOND} \\
    BTC   & 1.0000 & \cellcolor[rgb]{ .741,  .843,  .933} 0.2535 & \cellcolor[rgb]{ .741,  .843,  .933} 0.3161 & \cellcolor[rgb]{ .741,  .843,  .933} 0.1912 & \cellcolor[rgb]{ .886,  .937,  .855} -0.0114 & \cellcolor[rgb]{ .886,  .937,  .855} 0.0356 & \cellcolor[rgb]{ .886,  .937,  .855} -0.0371 \\
    DASH  &       & 1.0000 & \cellcolor[rgb]{ .741,  .843,  .933} 0.2072 & \cellcolor[rgb]{ .741,  .843,  .933} 0.2035 & \cellcolor[rgb]{ .886,  .937,  .855} -0.0124 & \cellcolor[rgb]{ .886,  .937,  .855} 0.0190 & \cellcolor[rgb]{ .886,  .937,  .855} -0.0574 \\
    XMR   &       &       & 1.0000 & \cellcolor[rgb]{ .741,  .843,  .933} 0.1639 & \cellcolor[rgb]{ .886,  .937,  .855} 0.0064 & \cellcolor[rgb]{ .886,  .937,  .855} 0.0004 & \cellcolor[rgb]{ .886,  .937,  .855} -0.0699 \\
    XRP   &       &       &       & 1.0000 & \cellcolor[rgb]{ .886,  .937,  .855} 0.0171 & \cellcolor[rgb]{ .886,  .937,  .855} 0.0029 & \cellcolor[rgb]{ .886,  .937,  .855} 0.0411 \\
    GOLD  &       &       &       &       & 1.0000 & \cellcolor[rgb]{ .973,  .796,  .678} -0.1023 & \cellcolor[rgb]{ .973,  .796,  .678} 0.2815 \\
    S\&P500 &       &       &       &       &       & 1.0000 & \cellcolor[rgb]{ .973,  .796,  .678} -0.3466 \\
    BOND  &       &       &       &       &       &       & 1.0000 \\
    \end{tabular}%
		}
  \label{tab:matrix}%
\end{table}%

\section{Conclusions \label{sec:conclusions}}
Our empirical study detects that cryptocurrencies exhibit similar mean correlation among them ($\approx 0.20$), with a maximum of $0.31$ and a minimum of $0.16$. However, these correlations vary across time. The standard deviation of the pairs including Monero is almost half the standard deviation of other pairs. As a result, those pairs generate more stable correlation across time. The other pairs exhibit large variations, from $\approx-0.50$ to $\approx+0.70$. 

We also compute correlation against more traditional assets, such as stock and bond indices and gold. In this case we detect that the cryptocurrency market is detached from the behavior of other financial markets. In addition, we find a singular behavior of Monero \textit{vis-\`a-vis} the traditional assets: cross correlation is more stable across time, and closer to zero, compared to other cryptocurrencies.

\bibliographystyle{apalike}
\bibliography{longmemorybib}

\begin{thebibliography}{}

\bibitem[Aslanidis and Casas, 2013]{Aslanidis2013}
Aslanidis, N. and Casas, I. (2013).
\newblock {Nonparametric correlation models for portfolio allocation}.
\newblock {\em Journal of Banking {\&} Finance}, 37(7):2268--2283.

\bibitem[Balcilar et~al., 2017]{Balcilar20177em}
Balcilar, M., Bouri, E., Gupta, R., and Roubaud, D. (2017).
\newblock Can volume predict bitcoin returns and volatility? a quantiles-based
  approach.
\newblock {\em Economic Modelling}, 64:74 -- 81.

\bibitem[Bariviera, 2017]{Bariviera2017}
Bariviera, A.~F. (2017).
\newblock {The inefficiency of Bitcoin revisited: A dynamic approach}.
\newblock {\em Economics Letters}, 161:1--4.

\bibitem[Bariviera et~al., 2018]{Bariviera2018}
Bariviera, A.~F., Zunino, L., and Rosso, O.~A. (2018).
\newblock {An analysis of high-frequency cryptocurrencies prices dynamics using
  permutation-information-theory quantifiers}.
\newblock {\em Chaos: An Interdisciplinary Journal of Nonlinear Science},
  28(7):075511.

\bibitem[Bouri et~al., 2017a]{Bourietal2017}
Bouri, E., Azzi, G., and Dyhrberg, A.~H. (2017a).
\newblock On the return-volatility relationship in the bitcoin market around
  the price crash of 2013.
\newblock {\em Economics: The Open-Access, Open-Assessment E-Journal},
  11:1--16.

\bibitem[Bouri et~al., 2017b]{Bouri20171frl}
Bouri, E., Molnár, P., Azzi, G., Roubaud, D., and Hagfors, L.~I. (2017b).
\newblock On the hedge and safe haven properties of bitcoin: Is it really more
  than a diversifier?
\newblock {\em Finance Research Letters}, 20:192 -- 198.

\bibitem[Cappiello et~al., 2006]{Cappiello2006}
Cappiello, L., Engle, R.~F., and Sheppard, K. (2006).
\newblock {Asymmetric Dynamics in the Correlations of Global Equity and Bond
  Returns}.
\newblock {\em Journal of Financial Econometrics}, 4(4):537--572.

\bibitem[Coinmarket, 2018]{coinmarketcap}
Coinmarket (2018).
\newblock {Crypto-Currency Market Capitalizations}.
\newblock \url{https://coinmarketcap.com/currencies/}.
\newblock Accessed: 2018-10-01.

\bibitem[Corbet et~al., 2018a]{Corbet2018b}
Corbet, S., Lucey, B., Urquhart, A., and Yarovaya, L. (2018a).
\newblock {Cryptocurrencies as a financial asset: A systematic analysis}.
\newblock {\em International Review of Financial Analysis}.

\bibitem[Corbet et~al., 2018b]{Corbet2018}
Corbet, S., Meegan, A., Larkin, C., Lucey, B., and Yarovaya, L. (2018b).
\newblock {Exploring the dynamic relationships between cryptocurrencies and
  other financial assets}.
\newblock {\em Economics Letters}, 165:28--34.

\bibitem[Diebold and Yilmaz, 2012]{Diebold2012}
Diebold, F.~X. and Yilmaz, K. (2012).
\newblock {Better to give than to receive: Predictive directional measurement
  of volatility spillovers}.
\newblock {\em International Journal of Forecasting}, 28(1):57--66.

\bibitem[Donier and Bouchaud, 2015]{DonierBouchaudPlosOne}
Donier, J. and Bouchaud, J.-P. (2015).
\newblock Why do markets crash? bitcoin data offers unprecedented insights.
\newblock {\em PLOS ONE}, 10(10):1--11.

\bibitem[Dyhrberg, 2016]{Dyhrberg2016}
Dyhrberg, A.~H. (2016).
\newblock {Bitcoin, gold and the dollar ? A GARCH volatility analysis}.
\newblock {\em Finance Research Letters}, 16:85--92.

\bibitem[Elbahrawy et~al., 2017]{Elbahrawy2017}
Elbahrawy, A., Alessandretti, L., Kandler, A., Pastor-Satorras, R., and
  Baronchelli, A. (2017).
\newblock {Evolutionary dynamics of the cryptocurrency market}.
\newblock {\em Royal Society Open Science}, 4(11).

\bibitem[Engle, 2002]{Engle2002}
Engle, R. (2002).
\newblock {Dynamic Conditional Correlation}.
\newblock {\em Journal of Business {\&} Economic Statistics}, 20(3):339--350.

\bibitem[Hafner et~al., 2006]{Hafner2006}
Hafner, C.~M., van Dijk, D., and Franses, P.~H. (2006).
\newblock {Semi-Parametric Modelling of Correlation Dynamics}.
\newblock In Terrell, D. and Fomby, T.~B., editors, {\em Econometric Analysis
  of Financial and Economic Time Series (Advances in Econometrics}, pages
  59--103. Emerald Group Publishing Limited.

\bibitem[Kristoufek, 2015]{Kristoufek2015}
Kristoufek, L. (2015).
\newblock {What Are the Main Drivers of the Bitcoin Price? Evidence from
  Wavelet Coherence Analysis}.
\newblock {\em PLOS ONE}, 10(4):e0123923.

\bibitem[Nadarajah and Chu, 2017]{Nadarajah2017}
Nadarajah, S. and Chu, J. (2017).
\newblock {On the inefficiency of Bitcoin}.
\newblock {\em Economics Letters}, 150:6--9.

\bibitem[Nakamoto, 2009]{Nakamoto}
Nakamoto, S. (2009).
\newblock Bitcoin: A peer-to-peer electronic cash system.
\newblock \url{https://bitcoin.org/bitcoin.pdf/}.
\newblock Accessed: 2016-12-27.

\bibitem[Polasik et~al., 2015]{Polasik2015}
Polasik, M., Piotrowska, A.~I., Wisniewski, T.~P., Kotkowski, R., and
  Lightfoot, G. (2015).
\newblock {Price Fluctuations and the Use of Bitcoin: An Empirical Inquiry}.
\newblock {\em International Journal of Electronic Commerce}, 20(1):9--49.

\bibitem[Smith and Kumar, 2018]{SmithKumar}
Smith, C. and Kumar, A. (2018).
\newblock Crypto-currencies - an introduction to not-so-funny moneys.
\newblock {\em Journal of Economic Surveys}, 0(0).

\bibitem[Urquhart, 2016]{Urquhart2016}
Urquhart, A. (2016).
\newblock {The inefficiency of Bitcoin}.
\newblock {\em Economics Letters}, 148:80--82.

\end{thebibliography}

\end{document}